\documentclass[aps,prb,twocolumn,showpacs,floatfix,superscriptaddress]{revtex4}
\usepackage{dcolumn}
\usepackage{bm}
\usepackage{amsmath,amssymb,graphicx}

\begin{document}
\preprint{Submitted to Phys. Rev. B}

\title{Circular-Polarization Dependent Cyclotron Resonance in Large-Area Graphene in Ultrahigh Magnetic Fields}

\author{L.~G.~Booshehri}
\affiliation{Department of Electrical and Computer Engineering, Rice University, Houston, Texas 77005}
\affiliation{National High Magnetic Field Laboratory, Los Alamos, New Mexico 87545}

\author{C.~H.~Mielke}
\author{D.~G.~Rickel}
\author{S.~A.~Crooker}
\affiliation{National High Magnetic Field Laboratory, Los Alamos, New Mexico 87545}

\author{Q.~Zhang}
\author{L.~Ren}
\author{E.~H.~H\'{a}roz}
\affiliation{Department of Electrical and Computer Engineering, Rice University, Houston, Texas 77005}

\author{A.~Rustagi}
\author{C.~J.~Stanton}
\affiliation{Department of Physics, University of Florida, Gainesville, Florida, 32611}

\author{Z.~Jin}
\author{Z.~Sun}
\author{Z.~Yan}
\affiliation{Department of Chemistry, Rice University, Houston, Texas 77005}

\author{J.~M.~Tour}
\affiliation{Department of Chemistry, Rice University, Houston, Texas 77005}
\affiliation{Department of Computer Science, Rice University, Houston, Texas 77005}
\affiliation{Department of Mechanical Engineering and Materials Science, Rice University, Houston, Texas 77005}

\author{J.~Kono}
\email[]{kono@rice.edu}
\homepage[]{www.ece.rice.edu/~kono}
\thanks{corresponding author.}
\affiliation{Department of Electrical and Computer Engineering, Rice University, Houston, Texas 77005}
\affiliation{Department of Physics and Astronomy, Rice University, Houston, Texas 77005}

\date{\today}
\begin{abstract}
Using ultrahigh magnetic fields up to 170~T and polarized midinfrared radiation with tunable wavelengths from 9.22 to 10.67~$\mu$m, we studied cyclotron resonance in large-area graphene grown by chemical vapor deposition.  Circular-polarization dependent studies reveal strong $p$-type doping for as-grown graphene, and the dependence of the cyclotron resonance on radiation wavelength allows for a determination of the Fermi energy.  Thermal annealing shifts the Fermi energy to near the Dirac point, resulting in the simultaneous appearance of hole {\em and} electron cyclotron resonance in the magnetic quantum limit, even though the sample is still $p$-type, due to graphene's linear dispersion and unique Landau level structure.  These high-field studies therefore allow for a clear identification of cyclotron resonance features in large-area, low-mobility graphene samples.
\end{abstract}

\maketitle


The band structure of graphene exhibits a zero-gap linear dispersion relation near each of the Dirac points, which results in a variety of exotic properties of two-dimensional (2D) Dirac fermions.\cite{Novoselov:2005p18661, Zhang:2005p18662, Katsnelson:2006p18663}  While a number of electronic transport studies have revealed novel phenomena in the presence of a high magnetic field, including half-integer quantum Hall states observed at room temperature,\cite{Novoselov:2005p18661, Zhang:2005p18662, Zhang:2006p18664} magneto-optical properties are expected to be equally unusual,\cite{ZhengAndo02PRB,AndoetAl02JPSJ,GusyninSharapov06PRB,GusyninetAl06PRL,IyengaretAl07PRB,Abergel:2007p18682,PereiraetAl07PRB,Gusynin:2007p18665,KoshinoAndo08PRB,BarlasetAl08PRL,Mikhailov:2009p18666}  especially in the magnetic quantum limit\cite{Gusynin:2007p18665} where the Fermi level resides in the lowest Landau level (LL).  Even in conventional 2D electron systems such as found in GaAs quantum wells, studies of cyclotron resonance (CR) in the magnetic quantum limit have shown many-body effects,\cite{Besson:1992p18667, Summers:1993p18668, Cooper:1994p18669, Asano:1996p18670} such as spin-splitting in the fractional quantum Hall regime, even though CR is not expected to be sensitive to electron-electron interactions due to Kohn's theorem.\cite{Kohn:1961p18671}  The linear dispersions of graphene automatically evade this basic requirement for Kohn's theorem, motivating CR studies of graphene in ultrahigh magnetic fields.

An applied magnetic field ($B$) creates LLs for charge carriers both in the conduction and valence bands, and CR measures resonant optical transitions between adjacent LLs ($\Delta n$ = $\pm$1, where $n$ is the Landau level index).\cite{Kono01MMR} CR is a well-established and powerful technique to determine many fundamental parameters of a sample, such as carrier effective masses, densities, mobilities and scattering rates.  When performed with circularly polarized radiation, the sign of the charge carriers can also be determined.  Furthermore, owing to graphene's nonparabolic (i.e., linear) dispersion, LL energies are not equally spaced; rather, they follow $E_{n,\pm}=\pm c^{*} \sqrt{2e\hbar Bn}$, where $n \geq$ 0 and $c^{*}$ = 1.0 $\times$ 10$^{6}$~m/s corresponds to the slope of the linear dispersions.  Thus, different inter-Landau level (LL) transitions occur at \textit{different} energies or magnetic fields.  Hence, the absence or presence of a certain resonance can determine the Fermi energy.  This is in marked contrast to conventional materials with parabolic dispersions, which form equally spaced LLs in a magnetic field ($E_n = (n + 1/2)e\hbar B /m^{*}$, where $m^{*}$ is the effective mass) and in which all inter-LL transitions ($\Delta n$ = $\pm$ 1) occur at the same energies (for a given magnetic field) or at the same magnetic field (for a given probe photon energy).

A number of CR measurements have been performed on graphene,\cite{Sadowski:2006p18674,Jiang:2007p18672,Deacon:2007p18673,HenriksenetAl08PRL,OrlitaetAl08PRL,NeugebaueretAl09PRL,HenriksenetAl10PRL,CrasseeetAl11NP,Witowski:2010p18675,OrlitaetAl11PRL,CrasseeetAl11PRB} successfully resolving the unusual LL structure, particularly when the graphene samples investigated have relatively high mobilities, such as exfoliated graphene or epitaxial graphene on SiC.  However, for technologically important applications requiring large-area graphene films grown via chemical vapor deposition (CVD), we still face the current problem of low mobilities ($\sim$10$^{3}$~cm$^{2}$V$^{-1}$s$^{-1}$), which severely broadens CR.  Therefore, for these low-mobility samples, CR measurements must be explored within the regime of ultrahigh magnetic fields that makes the CR observability condition $\omega_{c}\tau >$ 1 achievable (here, $\omega_{c}$ is the cyclotron frequency and $\tau$ is the carrier scattering time).  Furthermore, a pulsed high-field magnet can be readily combined with an infrared laser to perform CR measurements with circularly-polarized radiation,\cite{Kono01MMR, KonoMiura06HMF} allowing us to distinguish between electron-like and hole-like CR transitions.  Magneto-spectroscopy using circularly-polarized radiation indeed played a crucial role in elucidating electron and hole states in graphite.\cite{SchroederetAl68PRL,SeumatsuTanuma72JPSJ,ToyetAl77PRB,DoezemaetAl79PRB}  Furthermore, a recent magneto-optical study on multilayer graphene on SiC using circularly-polarized radiation revealed multiple CR components, demonstrating the existence of multiple species of carriers with different Fermi velocities.\cite{CrasseeetAl11PRB}

Here, we present an experimental study of CR in CVD-grown graphene at ultrahigh magnetic fields.  Circularly-polarized magneto-transmission of CO$_{2}$ laser light (wavelength $\lambda$ = 9.22 $\mu$m to 10.7 $\mu$m) revealed a strong and unintentional $p$-type (hole) doping of these nominally-undoped graphene samples.  A small CR feature at 10 T and a larger CR feature at 65 T corresponds to the $n$ = 0 to $n = -1$ and $n = -1$ to $n = -2$ inter-LL transitions, respectively.  Wavelength-dependent magneto-transmission data show a good agreement with calculated LL transitions, allowing the Fermi energy ($E_{F}$) of the sample to be determined ($-$250 meV).  Moreover, we found that annealing the graphene samples in vacuum to remove physisorbed molecules shifts the Fermi energy markedly closer to the Dirac point ($E_{F}$ = $-$34 meV).  Due to graphene's unique Landau level structure, we show that in the magnetic quantum limit this allows both hole ($n$ = 0 to $n = -1$) {\em and} electron ($n$ = 0 to $n = +1$) CR to be observed {\em simultaneously}, even though the sample is still $p$-type. 


We measured graphene samples grown via CVD on a large-area ($\sim$4 cm$^{2}$) copper foil, resulting in the growth of large-area, high-quality graphene.\cite{Jin:2011p18676}  A poly (methyl methacrylate) (PMMA) film was spin-coated onto the graphene-covered copper foil and was heated up to 60 $^{\circ}$C for 5 min to cure the PMMA film.  Then the PMMA/ graphene layer was separated from the copper foil by etching in a 1 M CuCl$_{2}$/6 M HCl aqueous solution and was placed on the surface of Milli-Q water to remove any water-soluble residues.  The PMMA/ graphene film was then transferred onto a thallium bromoiodide (KRS-5) substrate, which is transparent at mid-infrared wavelengths.  The PMMA film was dissolved away by soaking the substrate in acetone for 8 hours,\cite{Jin:2011p18676} thus leaving only the CVD-graphene sheets remaining on the KRS-5 substrate.

\begin{figure}
\includegraphics[scale=0.6]{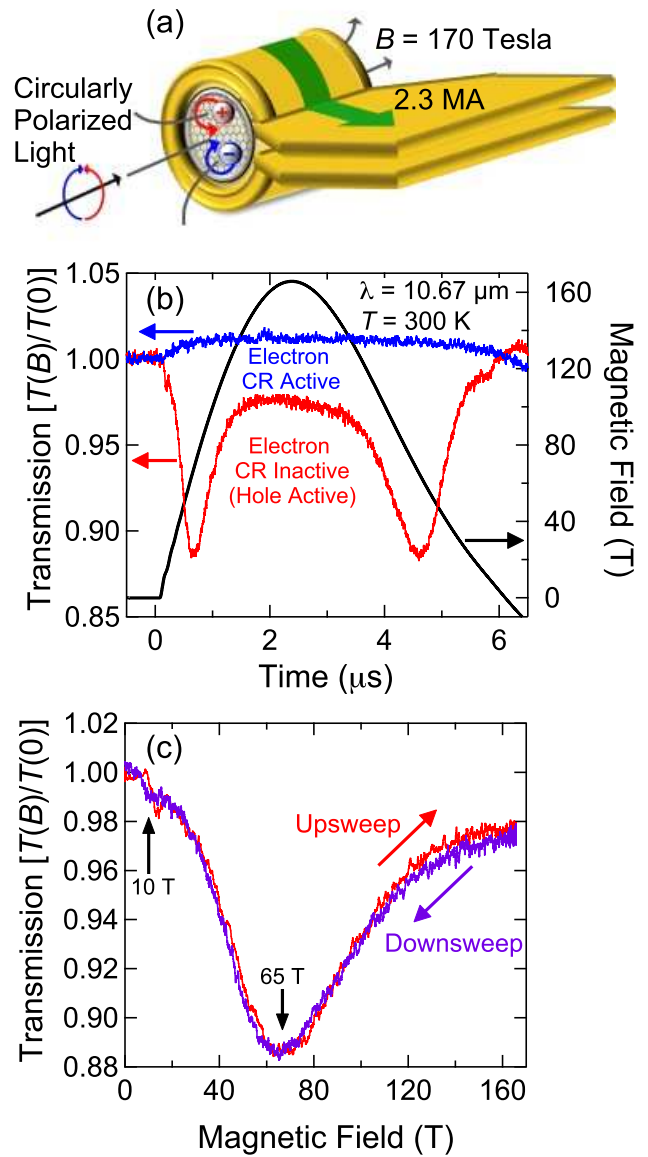}
\caption{\label{fig1}(color online) (a) Experimental configuration for magneto-transmission using the STC magnet.  (b)~Magneto-transmission of 10.67~$\mu$m light through a nominally-undoped CVD-grown graphene sample, during a 170~T magnet pulse.  Data for both circular polarizations are shown.  The pronounced CR absorption that appears for ``electron CR \textit{inactive}" polarization indicates that the graphene is $p$-type (hole doped).  (c)~Electron CR inactive transmission versus magnetic field shows two CR features at 10~T and 65~T, corresponding to $n$ = 0 to $n = -1$ and $n = -1$ to $n = -2$ inter-LL transitions.}
\end{figure}

Transmission measurements were performed at room temperature in the Faraday geometry (where light propagates parallel to the magnetic field), using a continuous-wave, single-mode CO$_{2}$ laser that is tunable from 9.2 to 10.7~$\mu$m.  The combination of a linear polarizer and quarter-wave plate was used to switch between opposite circular polarizations (historically called ``electron CR active" and ``electron CR inactive"), and the transmission was detected with a mercury cadmium telluride detector sensitive in the 2-12~$\mu$m spectral range.  The handedness of the circular polarization and direction of the magnetic field were cross-checked via electron CR of bulk $n$-type GaAs.

To access ultrahigh magnetic fields, we used the Single Turn Coil (STC) magnet at the National High Magnetic Field Laboratory (NHMFL) in Los Alamos\cite{Mielke:2006p18677, Mielke:2010p18678} (see Fig.~\ref{fig1}a).  The STC magnet combines a low-inductance, 60-kV capacitor bank with a 10-mm diameter, 10-mm long single-turn copper coil that discharges a maximum current of $\sim$4~MA in $\sim$6~$\mu$s and reaches magnetic fields in excess of 200~T during standard operation.  The large current, magnetic field, and associated Lorentz forces cause an outward, radial expansion (and eventual destruction) of the copper coil within $\sim$15~$\mu$s.  However, the sample and sample probe are not damaged, and repeated measurements on the sample are possible.


Figure~\ref{fig1}(b) shows the magneto-transmission of both electron CR active and electron CR inactive circularly polarized light at 10.67~$\mu$m through a nominally-undoped graphene sample at room temperature.  The magnetic field reaches a peak field of 170~T in less than 2.5~$\mu$s, with a total rise and fall time of $\sim$6~$\mu$s.  Within this time duration, we observe clear CR absorption as transmission dips only for the electron CR {\em inactive} polarization.  Therefore, not only does this graphene sample have a nonzero density of charge carriers, these charge carriers are {\em holes}.  This $p$-type doping is typical for as-grown, nominally-undoped graphene, most likely due to physisorption of molecules, such as H$_{2}$O and O$_{2}$, that causes the Fermi energy of undoped graphene to be shifted away from the Dirac point and into the valence band.\cite{Jin:2011p18676, Schedin:2007p18679}

The data actually reveal {\em two} CR features, one large absorption at $\sim$65~T and a smaller absorption at $\sim$10~T [Fig.~\ref{fig1}(c)].  We also notice the lack of any hysteresis of the transmission data between the upsweep and the downsweep of the magnetic field, which not only rules out any heating effects but also verifies the sufficient speed of the measurement system.  From the linewidth of the larger CR feature, we can calculate the effective carrier scattering time, $\tau$, to be $\sim$14~fs ($\omega_{c}\tau$ = $B_{c}/\Delta B_{c}$), which is typical for CVD-grown graphene.\cite{Tan:2007p18680}

\begin{figure} [tbp]
\includegraphics[scale=0.61]{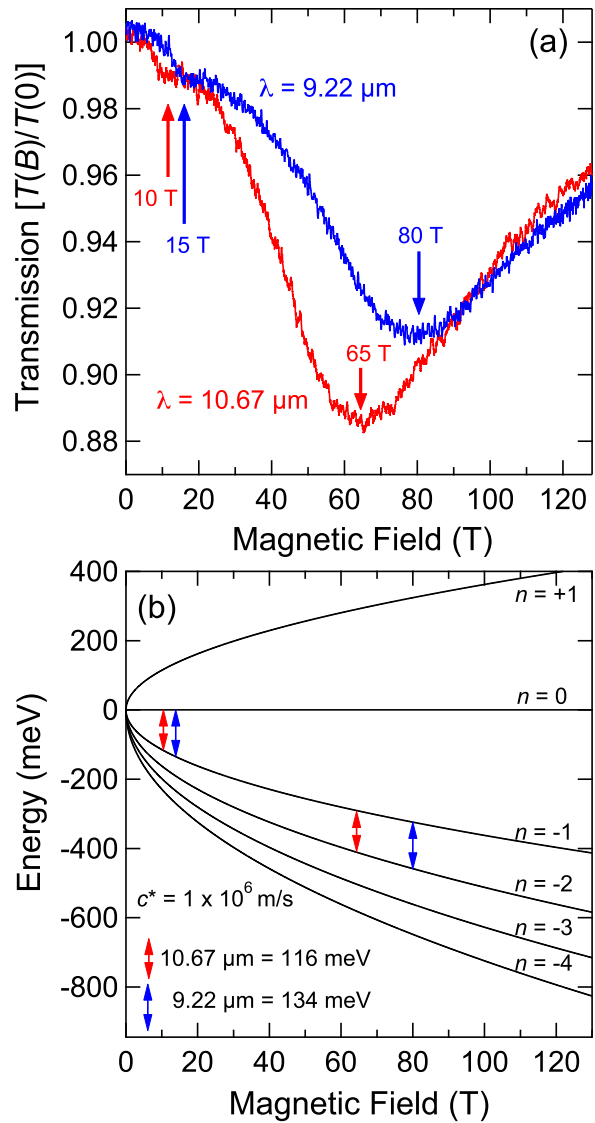}
\caption{\label{fig2}(color online) (a) Wavelength-dependent electron-CR-inactive transmission traces versus magnetic field.  Each trace exhibits two CR features (hole CR).  (b)~Landau level fan diagram with calculated LL
transitions for 10.67~$\mu$m (red) and 9.22~$\mu$m (blue).}
\end{figure}

Figure~\ref{fig2}(a) compares magneto-transmission data using 9.22~$\mu$m and 10.67~$\mu$m radiation (134~meV and 116~meV photons, respectively).  We again observe two CR absorptions for each wavelength, and we find that the resonance fields increase from 10 to 15~T and 65 to 80~T, respectively, with increasing photon energy.  In the Appendix, we calculate the Landau levels and the expressions for circularly polarized CR absorption.  Using a Landau level fan diagram calculated for graphene, Fig.~\ref{fig2}(b) shows the calculated $n$ = 0 to $n = -1$ and $n = -1$ to $n = -2$ transitions at 9.22~$\mu$m (blue) and 10.67~$\mu$m (red).  Comparing our measured CR features to the calculated inter-LL transition energies, we can assign the low-field (high-field) feature to the $n$ = 0 to $n = -1$ ($n = -1$ to $n = -2$) transition.  

Moreover, taking into account the measured integrated peak intensities from Fig.~\ref{fig2}(a), we see a decrease in CR absorption with increasing photon energy due to a decrease in population of the $n = -1$ LL with increasing magnetic field.  Because we do not observe any CR absorption corresponding to the $n = -2$ to $n = -3$ transition (expected at 103~T and 135~T for these wavelengths, respectively), we can first conclude that the Fermi energy $E_{F}$ must reside within the $n = -1$ LL at 65~T and must therefore have a zero-field value in the range between $-$140 to $-$400~meV.  However, to more accurately determine the Fermi energy, we need to consider the degeneracy of each LL, which is $eB/2\pi\hbar$ ($\times$ 4 for valley and spin degeneracy), at the resonance fields 65~T and 80~T.  Calculating the ratio of the 65-T CR peak intensity to the 80-T CR peak intensity and using this ratio alongside the LL degeneracy, we can estimate the hole occupation within the $n = -1$ LL at 65~T and 80~T.  We directly calculate the hole density of our sample, $n_{h}$ = 1.6 $\times$ 10$^{12}$~cm$^{-2}$, which translates to a zero-field Fermi energy, $E_{F}$ = $\hbar c^{*}\sqrt{4\pi n_{h}} \approx$ $-$295~meV.

\begin{figure*}
\includegraphics[scale=0.65]{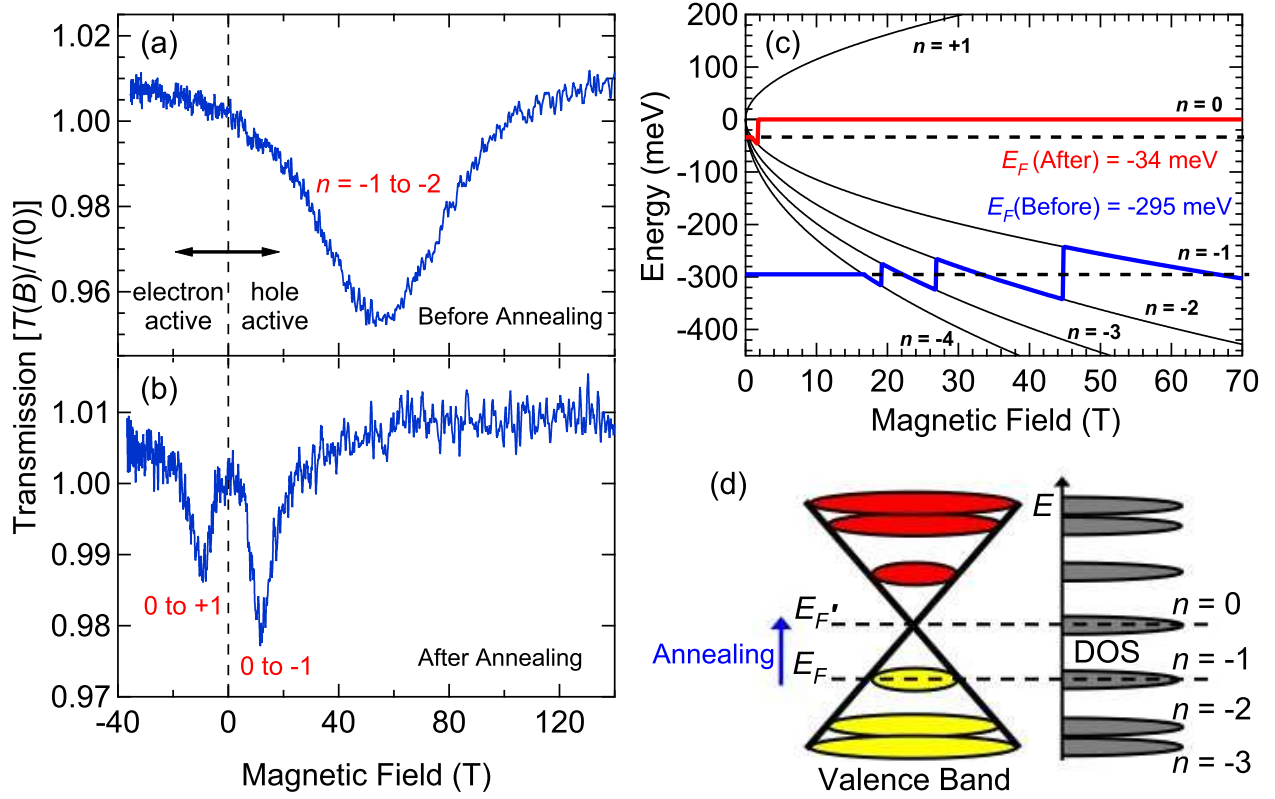}
\caption{\label{fig3}(color online) (a) Electron CR inactive transmission at 10.6~$\mu$m before annealing, and (b)~after annealing.  (c)~Landau fan diagram showing the Fermi energy oscillation with magnetic field for before (blue) and after (red) annealing, and (d)~depicts that annealing moves the Fermi energy from $-$295 to $-$34~meV.}
\end{figure*}

This relatively large value of $E_{F}$ due to accidental $p$-doping through physisorption of H$_{2}$O and O$_{2}$ is typical for graphene but can be controlled if we can remove the physisorbed molecules.  Figure~\ref{fig3} demonstrates that one can use thermal annealing to control the Fermi energy.  Here, we measured CR with electron CR inactive (that is, {\em hole} CR active) polarization before annealing [Fig.~\ref{fig3}(a)] and immediately after we annealed the sample for 2 hours at 150$^{\circ}$C and 10$^{-6}$~mbar to remove any physisorbed molecules [Fig.~\ref{fig3}(b)].  As shown in Fig.~\ref{fig3}(b), the majority of the doped holes are removed during the annealing process, and we now see two smaller CR peaks at $+$10~T and $-$10~T, corresponding to the $n$ = 0 to $n = -1$ transition and the $n$ = 0 to $n$ = +1 transition, i.e., to hole {\em and} electron CR in the magnetic quantum limit.  Note that, since the polarization is hole CR active for positive magnetic field, any resonance on the negative magnetic field side is due to electron CR (note that the field swings negative to about $-$40~T following a $+$170~T pulse, allowing both positive and negative fields to be examined in a single pulse).  The emergence of the two 10~T peaks with opposite signs, and the absence of the $\sim$60~T peak (the $n = -1$ to $n = -2$ transition), unambiguously demonstrate that the Fermi energy now resides in the $n$ = 0 LL --- the magnetic quantum limit --- at 10 T.

Furthermore, as the integrated hole CR absorption is stronger than the electron CR absorption [see Fig.~\ref{fig3}(b)], the graphene sample is still slightly $p$-type and the intensity ratio between the two peaks provides information about the ratio of occupied versus unoccupied states within the $n$ = 0 LL.  More specifically, comparing the peak intensities, we calculate the ratio of occupied to unoccupied states within the $n$ = 0 LL to be $\sim$1:1.5.  We can then use this ratio and the LL degeneracy to calculate the hole density of our sample after annealing to be $n_{h}$ = 2.1 $\times$ 10$^{10}$~cm$^{-2}$.  As a result, we can determine the zero-field Fermi energy after annealing to be $E_{F}$ = $-$34~meV, again through $E_{F}$ = $\hbar c^{*}\sqrt{4\pi n_{h}}$.  Figure~\ref{fig3}(c) shows the oscillations of the Fermi energy with increasing magnetic field for both before (blue) and after (red) annealing.  As Fig.~\ref{fig3}(d) depicts, before annealing the Fermi level sits within the $n = -1$ LL at 65~T and after annealing the Fermi energy now sits within the $n$ = 0 LL at 10~T.  The annealing process has successfully removed most of the physisorbed molecules that $p$-doped the graphene, leaving the system only slightly $p$-type.  Qualitatively similar thermal effects were observed for other monolayer graphene samples that we investigated.

It should be emphasized here that the simultaneous appearance of both hole {\em and} electron CR is an unusual phenomenon, made possible only via the unique electronic structure and Landau quantization of graphene.  This phenomenon cannot occur in conventional materials containing only one carrier type.  At $B$ = 0, our graphene samples are still slightly $p$-type even after annealing, and they contain {\em only hole carriers}.  However, as soon as a finite magnetic field is applied, a Landau level with exactly zero energy appears, i.e., the $n$ = 0 Landau level (4-fold degenerate due to valley and spin).  This level never moves with $B$ in energy but simply grows in degeneracy.  As long as the Fermi energy lies within this level, this level is {\em partially filled}, containing both holes {\em and} electrons, allowing the observation of {\em both} electron and hole CR even at $T$ = 0.

Detailed calculations confirm these results.  In the Appendix, we compute the CR absorption for both electron-active and hole-active circular polarizations.  The results are given by Eqs.~(\ref{electron LL}) and (\ref{negative electron LL}).  The results show a number of interesting and unusual features that we have discussed: i) The optical field Hamiltonian for graphene comes in through an $\bf{A}\cdot\bm{\sigma}$ term rather than the usual term $\bf{A}\cdot\bf{p}$; ii) the expressions show that even an {\em undoped} sample will show CR in contrast to conventional semiconductors where the sample must be doped; iii) if the Fermi energy lies in the $n$ = 0 level, then CR is present for $both$ electron-active and hole-active circular polarizations; and iv) if the Fermi level lies in the $n$ = 0 level, the ratio of CR absorption between the electron-active and hole-active peaks allows one to accurately determine the Fermi level and carrier density.

In summary, we have measured cyclotron resonance in graphene at ultrahigh magnetic fields.  From polarization and wavelength dependent measurements, we have shown unintentional chemical doping is strongly $p$-type by observing the $n$ = 0 to $n = -1$ and $n = -1$ to $n = -2$ transitions in nominally-undoped graphene.  Additionally, we demonstrated that annealing effectively removes $p$-doping from graphene, shifting the Fermi energy much closer to the Dirac point.  This shift completely blocked the $n = -1$ to $n = -2$ hole cyclotron resonance  transition and revealed the $n$ = 0 to $n = +1$ electron cyclotron resonance transition.  As a result, due to graphene's unique band structure, we simultaneously observed both hole and electron cyclotron resonance in the magnetic quantum limit, even with a $p$-type graphene sample.  These high-field studies therefore allow for a clear identification of cyclotron resonance features in large-area graphene samples, and an accurate determination of their low carrier mobilities and fast scattering times.

\begin{acknowledgments}
This work was supported by the National Science Foundation through Grant Nos.~DMR-1006663, DMR-1105437, and OISE-0968405, the Department of Energy BES Program through Grant No.~DEFG02-06ER46308, and the Robert A.~Welch Foundation through Grant No.~C-1509.  Work carried out at the National High Magnetic Field Laboratory -- Pulsed Field Facility at Los Alamos National Laboratory was provided by funding from the National Science Foundation through Grant No.~DMR-0654118 and from the Department of Energy BES program, the State of Florida, and the Los Alamos National Laboratory through the LDRD-DR program. Finally, this research was also made with Government support under and awarded by DoD, Air Force Office of Scientific Research, National Defense Science and Engineering Graduate (NDSEG), 32 CFR 168a.  S.A.C. acknowledges support from the UCOP program on Carbon Nanostructures.
\end{acknowledgments}

\appendix*
\section{CALCULATION OF CYCLOTRON RESONANCE AND INTERBAND MAGNETO-ABSORPTION FOR GRAPHENE}

Here, we calculate the circular polarization dependence of the cyclotron resonance absorption of graphene in a magnetic field.  We use CGS units here as opposed to the text so that we can compare our expressions to others.  In addition, we calculate the interband magneto-absorption (i.e., the valence band to conduction band transitions). 

The Hamiltonian for graphene near the K point is given by
\begin{equation}
H=\hbar c^{*} {\bm \sigma}\cdot {\bf k} = \hbar c^{*} 
\begin{pmatrix}0&k_{x}-ik_{y}\\k_{x}+ik_{y}&0\end{pmatrix},
\label{Graphene Hamiltonian}
\end{equation}
where $c^{*}$ is the Fermi velocity (and slope of the linear dispersion), $\bm{\sigma}$ are the Pauli matrices and $\bf{k}$ is wavevector measured from the K point $\bf{k} = \bf{k_{\Gamma}} - \bf{K}$ ($\bf{k_{\Gamma}}$ is the wavevector measured from the $\Gamma$ point).  The energy eigenvalues for this Hamiltonian are
\begin{equation}
\varepsilon=\pm\hbar c^{*} \lvert \bf{k} \rvert. 
\label{Energy eigenvalues}
\end{equation}
A similar Hamiltonian $H=\hbar c^{*}{\bm \sigma}^{*}\cdot\bf{k'}$ exists at the $\bf{K'}$ point but we will focus only on the K point.  In the results for our final expressions, we simply add a factor of four to account for both the spin degeneracy and the $\bf{K,K'}$ degeneracy.

In the presence of an external field (DC magnetic or AC optical), one uses the Peierls substitution (minimal coupling) ${\bf k} \rightarrow {\bf k} + e {\bf A}/\hbar c$ (note the electron charge is  $-e$), where A is the vector potential, to yield
\begin{equation}
H=\hbar c^{*} {\bm \sigma} \cdot \left( {\bf k}+\frac{e{\bf A}}{\hbar c}\right).
\label{Peierls Substitution}
\end{equation}
To determine the cyclotron resonance and absorption, we will consider the case of both a DC magnetic field and AC optical field so that the Hamiltonian becomes
\begin{equation}
H=\hbar c^{*} {\bm \sigma} \cdot \left( {\bf k}+\frac{e {\bf A_{\it B}}}{\hbar c} +\frac{e{\bf A_{\rm opt}}}{\hbar c} \right).
\label{AC-DC field}
\end{equation}
The magnetic field will be oriented in the $\hat{z}$ direction (perpendicular to the graphene layer), and we will use the Landau gauge in the form
\begin{equation}
{\bf A}_B = (0,Bx,0).
\label{Landau Gauge}
\end{equation}
For the AC optical field, we will take plane waves propagating in the $\hat{z}$ direction (i.e., in the same direction as the magnetic field).  The vector potential ${\bf A}_{\rm opt}$ for the photon field can be related to the AC electric field by 
\begin{eqnarray}
{\bf E_{\rm opt}} = {\bf E}_0 \exp \left[ i \left( {\bf q}\cdot{\bf r} - \omega t \right) \right] \\
\label{vector1}
%
{\bf A_{\rm opt}}=\frac{c {\bf E}_0}{i \omega} \exp \left[ i \left( {\bf q} \cdot {\bf r} - \omega t \right)\right].
\label{vector1}
\end{eqnarray}
Here $\bf{q}$ is in the $\hat{z}$ direction and ${\bf E}_0$ can be left or right circularly polarized in the $x$-$y$ plane.  We write the Hamiltonian [Eq.~(\ref{AC-DC field})] as a sum of an unperturbed term and a perturbation term
\begin{equation}
H=H_{\textit{B}}+H^{1},
\label{Sum Hamiltonian}
\end{equation}
where $H_{\textit{B}}$ is the unperturbed Hamiltonian with only a DC magnetic field.  This can be solved exactly by replacing $\bf{k}$ with $\hbar\nabla$/$i$ in Eq.~(\ref{AC-DC field}) and keeping only the magnetic field term.  The spectrum in the absence of the perturbation $H^{1}$ is simply the Landau level spectra
\begin{equation}
E_{n,\pm}=\pm\gamma\sqrt{n},
\label{Landau spectra}
\end{equation}
where $\gamma=\frac{\sqrt{2}c^{*}\hbar}{l_{B}}=c^{*}\sqrt{\frac{2e\hbar B}{c}}$, $n=0,1,2...$, and the magnetic length, $l_{B}$, is given by $l_{B}=\sqrt{\hbar c/eB}$.  Note that the energies vary with $\sqrt{n}$ and also with $\sqrt{B}$ (through the magnetic length). Note also that there is an $n$ = 0 level.  We can also write Eq.~(\ref{Landau spectra}) in a simplified form
\begin{equation}
E_{n,\pm}=\pm\hbar\omega_{0}\sqrt{n}\:,
\label{Simplified Landau}
\end{equation}
with $\omega_{0}=\sqrt{2}c^{*}/l_{B}$ being the characteristic frequency.  The wavefunctions are two-component spinors given by
\begin{equation} 
\langle x,y \mid n,\pm,k \rangle=\Psi_{n,\pm}(x,y)=b_{n}\begin{pmatrix} \Phi_{n-1,k} \\ \pm i\:\Phi_{n,k}  \end{pmatrix}.
\label{Spinors}
\end{equation}
We treat $n$ as {\em positive} or {\em zero} and the $+$ sign corresponds to the positive energy solution and the $-$ sign corresponds to the negative energy solution.  The normalization is given by
\begin{equation} 
b_{n} =\left \{ \begin {array} {ll} 1&\mbox { if } 
n=0 \\ \frac{1}{\sqrt{2}} & \mbox { if } n\neq 0 \end{array}\right.\\
\label{Normalization}
\end{equation}
and the component functions $\Phi_{n}$ are given by 
\begin{eqnarray}
\label{Component functions}
\nonumber &&
\Phi_{n.k}=\frac{1}{\sqrt{2^{n}n!\sqrt{\pi}}}\frac{1}{\sqrt{L_{y}l_{B}}}e^{i\:k\:y}\exp\left(-\frac{(x-kl^{2}_{B})^{2}}{2l^2_{B}}\right)
\\ &&
\times H_{n}\left(\frac{(x-kl_{B}^{2})}{l_{B}}\right),
\end{eqnarray}
with $\Phi_{-1}$=0, $H_{n}$ being the Hermite Polynomial of order $n$, and $k$ the quantum number associated with the $y$ motion in the Landau gauge.  The wavefunctions in Eq.~(\ref{Component functions}) are harmonic oscillator wavefunctions in the $x$ direction and plane waves in the $y$ direction.

We will treat the radiative Hamiltonian $H^{1}$ perturbatively using the spinor wavefunctions in Eq.~(\ref{Spinors}).  The perturbation term is
\begin{eqnarray}
\label{Perturbation Term} 
\nonumber &&
H^{1}=c^{*}\bm{\sigma}\cdot\left(\frac{e{\bf A}_{\rm opt}}{c}\right)=-\frac{iec^{*}}{\omega}\exp(i\:\bf{q}\cdot\bf{r})
\\&&
\times \begin{pmatrix}0&E_{x}-iE_{y}\\E_{x}+iE_{y}&0\end{pmatrix}.
\end{eqnarray}
This is not the usual $\bf{A}\cdot\bf{p}$ term that one uses in calculating the optical properties of conventional semiconductors.  If we had used the entire Hamiltonian for graphene instead of the Dirac approximation for the $\bf{K}$ and $\bf{K^{'}}$  points we could have used an $\bf{A}\cdot\bf{p}$ term in treating this interaction.

The optical matrix element corresponding to a transition caused by the perturbation term (for in-plane polarization, i.e., $\bf{q}$ is along the $z$-direction) from state $n$ to $m$ is calculated to be
\begin{eqnarray}
\label{State n to m} 
\nonumber &&
\langle m,\alpha,k'  \mid H^{1} \mid n, \beta, k \rangle = -\iint d^{2} r \frac{i\: e\: c^{*}}{\omega}e^{(i\: q\:z)}b_{m}b_{n} \times
\\ \nonumber &&
(\Phi_{m-1,k'}^{*},-\alpha i \Phi_{m,k'}^{*})\begin{pmatrix}0&E_{x}-iE_{y}\\E_{x}+iE_{y}&0\end{pmatrix}\begin{pmatrix}\Phi_{n-1,k}\\ \beta i \Phi_{n,k}\end{pmatrix}.
\\
\end{eqnarray}
We note that $n$ and $m$ are $\geq$ 0 and that $\alpha, \beta = \pm$  depending on whether the state is the positive energy state (conduction band) or the negative energy state (valence band).  Cyclotron resonance absorption corresponds to $\alpha = \beta$ and the inter-band magneto-absorption to $\alpha \neq \beta$.

For positive circularly polarized light  $\oplus$, we have ${\bf E_0} = E_0 (\hat{x} + i \hat{y})/\sqrt{2}$ so that $E_{x}=E_0/\sqrt{2}$, $E_{y} = i E_0/\sqrt{2}$ and the matrix element becomes:
\begin{eqnarray}
\label{Positive Circular Matrix Element} 
\nonumber &&
\langle m,\alpha,k'  \mid H^{1} \mid n, \beta, k \rangle_{\oplus} = -\iint d^{2} r \frac{i\: e\: c^{*}}{\omega}e^{(i\: q\:z)}b_{m}b_{n}
\\&&
\times(\Phi_{m-1,k'}^{*},-\alpha i \Phi_{m,k'}^{*})\begin{pmatrix}\beta\: i \sqrt{2} E_0 \Phi_{n,k}\\ 0\end{pmatrix},\end{eqnarray}
which integrates to 
\begin{eqnarray}
\label{Positive Circular Matrix Element Integrated} 
\nonumber &&
\langle m,\alpha,k'  \mid H^{1} \mid n, \beta, k \rangle_{\oplus} =\beta \frac{\sqrt{2}E_0\:e\:c^{*}}{\omega}e^{(i\: q\:z)}
\\&&
\times\: b_{m}\:b_{n}\:\delta_{m-1,n}\:\delta_{k,k'}.
\end{eqnarray}
Using the Fermi Golden Rule, we can evaluate the transition rate $W_{nm}$ for an electron in state $n$ going to an unoccupied state $m$, $n\rightarrow m$
\begin{equation} 
W_{nm}=\frac{2\pi}{\hbar}\lvert\langle m,\alpha,k'  \mid H^{1} \mid n, \beta, k \rangle\rvert^{2}\delta(\hbar\omega-\Delta E_{n,\beta}^{m,\alpha}),
\label{Wnm}
\end{equation}
\\
where $\Delta E_{n,\beta}^{m,\alpha}=E_{m, \alpha} - E_{n,\beta}$. Taking into account the probability $f_{n,\beta}$ that the state ($n$,$\beta$) is occupied and ($1-f_{m,\alpha}$) that the state ($m$,$\alpha$) is unoccupied, the transition rate \textit{per unit volume} is
\begin{eqnarray}
\label{Transition rate per unit volume} 
\nonumber &&
R_{n\rightarrow m}=\left(\frac{1}{V}\right)\frac{2\pi}{\hbar}\lvert\langle m,\alpha,k'  \mid H^{1} \mid n, \beta, k \rangle\rvert^{2}
\\&&
\times\: f_{n,\beta}(1-f_{m,\alpha})\delta(\hbar\omega-\Delta E_{n,\beta}^{m,\alpha}).
\end{eqnarray}
We must also take into account the inverse process (through emission of a photon), $m\rightarrow n$.  The transition rate per unit volume is given by
\begin{eqnarray}
\label{Inverse Transition rate} 
\nonumber &&
R_{m\rightarrow n}=\left(\frac{1}{V}\right)\frac{2\pi}{\hbar}\lvert\langle n,\beta,k'  \mid H^{1} \mid m, \alpha, k \rangle\rvert^{2}
\\&&
\times\: f_{m,\alpha}(1-f_{n,\beta})\delta(-\hbar\omega+\Delta E^{m,\alpha}_{n,\beta}).
\end{eqnarray}
The total \textit{net} rate/volume is the difference between the two, 
\begin{eqnarray}
\label{Net rate volume} 
\nonumber &&
R_{nm}=R_{n\rightarrow m}-R_{m\rightarrow n}
\\ \nonumber &&
\quad\quad\;=\left(\frac{1}{V}\right)\frac{2\pi}{\hbar}\lvert\langle m,\alpha,k' \mid H^{1} \mid n, \beta, k \rangle\rvert^{2}
\\ \nonumber &&
\qquad\qquad\times\: (f_{n,\beta}-f_{m,\alpha})\delta(\hbar\omega-\Delta E_{n,\beta}^{m,\alpha}).
\\
\end{eqnarray}
For undoped systems at $T$= 0, $f_{m,-}$ = 1 and $f_{m,+}$ = 0 for $m > 0$, i.e., all conduction band states are empty and all valence bands states are occupied, while $f_{0,-} = f_{0,+} = 1/2$. 

For positive circular polarized light,
\begin{eqnarray}
\label{Positive Circular polarized light} 
\nonumber &&
R_{mn}^{\oplus}=\frac{4\pi e^{2}c^{*2}}{V\hbar\omega^2}E_{0}^{2}b_{m}^{2}b_{n}^{2}
\\ \nonumber&&
\times\:(f_{n,\beta}-f_{m,\alpha})\delta_{m-1,n}\delta_{k,k'}\delta(\hbar\omega-\Delta E_{n,\beta}^{m,\alpha}).
\\
\end{eqnarray}
The total transition rate $R^{\oplus}$ summed over all possible transitions is
\begin{eqnarray}
\label{Summed transition rate} 
\nonumber &&
R_{mn}^{\oplus}=\frac{4}{V}\sum_{\substack{m,n\\ \alpha, \beta}}\sum_{k,k'}\frac{4\pi e^{2}c^{*2}}{\hbar \omega^{2}}E_{0}^{2}b_{m}^{2}b_{n}^{2}\left[\delta_{m-1,n}\:\delta_{k,k'}\right]^{2}
\\ \nonumber &&
\qquad\qquad\quad\quad\times\: (f_{n,\beta}-f_{m,\alpha})\delta(\hbar\omega-\Delta E_{n,\beta}^{m,\alpha})
\\ \nonumber &&
\quad\quad\;=\frac{4}{V}\sum_{\substack{n\geq 0\\ \alpha, \beta}}\sum_{k}\frac{2\pi e^{2}c^{*2}}{\hbar \omega^{2}}E_{0}^{2}b_{n}^{2}
\\ \nonumber &&
\qquad\qquad\quad\quad\times\: (f_{n,\beta}-f_{n+1,\alpha})\delta(\hbar\omega-\Delta E_{n,\beta}^{n+1,\alpha}).
\\
\end{eqnarray}
The factor of 4 comes from the 2-fold spin degeneracy and the 2-fold valley degeneracy.  Since the Landau levels are shifted harmonic oscillators in the \textit{x} direction centered about
\begin{equation} 
x_{0}=kl_{B}^{2}
\label{shifted harmonic}
\end{equation}
and since the oscillator center should be within the sample, following Roth\cite{Roth:1959p18681}, we require that
\begin{equation} 
0 < x_{0} < L_{x}\Rightarrow0<k<L_{x}/l_{B}^{2}.
\label{oscillator center}
\end{equation}
This restricts the allowed values of \textit{k}.  Converting the sum over \textit{k} into an integral
\begin{equation} 
\sum_{k} \rightarrow \int_{0}^{k_{max}}\frac{L_{y}}{2\pi}dk=\frac{L_{y}L_{x}}{2\pi l_{B}^{2}},
\label{sum over k to integral}
\end{equation}
it follows that 
\begin{eqnarray}
\label{Summed transition rate with graphene} 
\nonumber &&
R^{\oplus}=\frac{4}{L_{z}}\sum_{\substack{n\geq 0\\ \alpha, \beta}}\frac{e^{2}c^{*2}}{\hbar \omega^{2}l_{B}^{2}}E_{0}^{2}b_{n}^{2}
\\&&
\times\: (f_{n,\beta}-f_{n+1,\alpha})\delta(\hbar\omega-\Delta E_{n,\beta}^{n+1,\alpha}),
\end{eqnarray}
where $L_{z}$ is the thickness of the graphene layer.  The total power loss per unit volume is
\begin{eqnarray}
\label{power loss} 
\nonumber &&
P_{L}^{\oplus}=R^{\oplus}\hbar\omega=\frac{4}{L_{z}}\sum_{\substack{n\geq 0\\ \alpha, \beta}}\frac{e^{2}c^{*2}}{ \omega l_{B}^{2}}E_{0}^{2}b_{n}^{2}
\\&&
\times\: (f_{n,\beta}-f_{n+1,\alpha})\delta(\hbar\omega-\Delta E_{n,\beta}^{n+1,\alpha}).
\end{eqnarray}
The absorption coefficient, $\alpha_{abs}$, is defined to be\cite{Chuang1995}
\begin{eqnarray}
\label{define absorption} 
\nonumber &&
\alpha_{abs}=\frac{\#\: \text{of photons absorbed/unit volume} \times \text{second}}{\#\: \text{of photons injected/unit area} \times \text{second}}
\\&&
\quad\quad\,=\frac{R}{S/\hbar\omega}=\frac{R\hbar\omega}{S},
\end{eqnarray}
where $S$ is the magnitude of the average Poynting vector:
\begin{equation} 
S=\lvert\langle \text{\bf{S}}(\bm{r},t) \rangle\rvert=\frac{n_{r}cE_0^{2}}{8\pi}.
\label{poynting vector}
\end{equation}
Therefore, for positive circularly polarized light, 
\begin{eqnarray}
\label{positive absorption} 
\nonumber &&
\alpha_{abs}^{\oplus}=4\sum_{\substack{n\geq 0\\ \alpha, \beta}}\frac{8\pi e^{2}c^{*2}}{L_{z} \omega l_{B}^{2}n_{r}c}b_{n}^{2}
\\&&
\times\: (f_{n,\beta}-f_{n+1,\alpha})\delta(\hbar\omega-\Delta E_{n,\beta}^{n+1,\alpha}).
\end{eqnarray}
We will set $L_{z}=1$ to get the \textit{dimensionless} absorption per graphene layer.  For cyclotron resonance, $\alpha = \beta$, and for positive circularly polarized light, the requirement that
\begin{equation} 
\hbar\omega=\Delta E_{n,\beta}^{n+1,\beta}=\gamma\left(\beta\sqrt{n+1}-\beta\sqrt{n}\right)>0
\label{Requirement}
\end{equation}
restricts the cyclotron resonance transitions to the electron ($\beta$ = +) Landau levels and yields:
\begin{eqnarray}
\label{electron LL} 
\nonumber &&
\alpha_{abs}^{\oplus}=\frac{16\pi e^{2}c^{*2}\hbar}{n_{r} c l_{B}^{2}}\sum_{n\geq 0}\frac{2b_{n}^{2}}{\gamma\left(\sqrt{n+1}-\sqrt{n}\right)}
\\&&
\times\: (f_{n,+}-f_{n+1,+})\delta(\hbar\omega-\Delta E_{n,+}^{n+1,+}).
\end{eqnarray}
If we account for scattering which results in linewidth broadening, we can replace the delta function by a Lorentzian:
\begin{eqnarray}
\label{Lorentzian} 
\nonumber &&
\delta(\hbar\omega-\Delta E_{n,\beta}^{n+1,\alpha})\rightarrow\frac{1}{\pi}\frac{\hbar\tau^{-1}}{(\hbar\omega-\Delta E_{n,\beta}^{n+1,\alpha})^2+(\hbar\tau^{-1})^{2}}
\\ \nonumber &&
\qquad\qquad=\frac{1}{\pi}\frac{\hbar\tau^{-1}}{(\hbar\omega-\gamma\alpha\sqrt{n+1}+\gamma\beta\sqrt{n})^2+(\hbar\tau^{-1})^{2}},
\\
\end{eqnarray}
yielding the expression:
\begin{eqnarray}
\label{electron LL yield} 
\nonumber &&
\alpha_{abs}^{\oplus}=\frac{16\pi e^{2}c^{*2}\hbar}{n_{r} c l_{B}^{2}\gamma^{2}}\sum_{n\geq 0}\frac{2b_{n}^{2}}{\left(\sqrt{n+1}-\sqrt{n}\right)}(f_{n,+}-f_{n+1,+})
\\&&
\times\:\frac{\frac{\gamma\tau}{\pi\hbar}}{\frac{\gamma^{2}\tau^{2}}{\hbar^{2}}\left(\frac{\hbar\omega}{\gamma}-\sqrt{n+1}+\sqrt{n}\right)^{2}+1}.
\end{eqnarray}
This expression agrees with the results of Abergel and Fal'ko\cite{Abergel:2007p18682} (up to spin and valley degeneracy), who obtained these results using the Keldysh technique.

From  Eq.~(\ref{electron LL}), we see that the strength of the transitions depend upon $\sqrt{B}$.  A factor of $B$ comes from the 1/$l_{B}^{2}$ term (which comes from the degeneracy of the Landau levels) and a factor of 1/$\sqrt{B}$ comes from the 1/$\omega$ dependence of Eq.(\ref{power loss}) at the $\omega$=$\omega_0\sqrt{n}$ resonance.

For cyclotron resonance ($\alpha=\beta$) with negatively circularly polarized light, the relation between energy of the photon and Landau levels
\begin{equation} 
\hbar\omega=\Delta E_{n,\beta}^{n-1,\beta}=\gamma\left(\beta\sqrt{n-1}-\beta\sqrt{n}\right)>0
\label{Requirement}
\end{equation}
requires that $\beta = -$, i.e., only valence band transitions are allowed.  One can repeat the calculation for the cyclotron resonance for negative circularly polarized light in a similar manner to obtain:
\begin{eqnarray}
\label{negative electron LL} 
\nonumber &&
\alpha_{abs}^{-}=\frac{16\pi e^{2}c^{*2}\hbar}{n_{r} c l_{B}^{2}}\sum_{n\geq 1}\frac{2b_{n-1}^{2}}{\gamma\left(\sqrt{n}-\sqrt{n-1}\right)}
\\&&
\times\: (f_{n,-}-f_{n-1,-})\delta(\hbar\omega-\Delta E_{n,-}^{n-1,-}),
\end{eqnarray}
To obtain the inter-band magneto-absorption (valence band to conduction band), we use $\alpha\neq\beta$ and worry only about absorption (not emission) to get:
\begin{eqnarray}
\label{plus} 
\nonumber &&
\alpha_{abs}^{\oplus}=\frac{16\pi e^{2}c^{*2}\hbar}{n_{r} c l_{B}^{2}}\sum_{n\geq 1}\frac{2b_{n}^{2}}{\gamma\left(\sqrt{n+1}+\sqrt{n}\right)}
\\&&
\times\: (f_{n,-}-f_{n+1,+})\delta(\hbar\omega-\Delta E_{n,-}^{n+1,+}),
\end{eqnarray}
and
\begin{eqnarray}
\label{minus} 
\nonumber &&
\alpha_{abs}^{-}=\frac{16\pi e^{2}c^{*2}\hbar}{n_{r} c l_{B}^{2}}\sum_{n\geq 1}\frac{2b_{n-1}^{2}}{\gamma\left(\sqrt{n}+\sqrt{n-1}\right)}
\\&&
\times\: (f_{n,-}-f_{n-1,+})\delta(\hbar\omega-\Delta E_{n,-}^{n-1,+}).
\end{eqnarray}
Note that the usual $\Delta n$ = 0 selection rule for the inter-band magneto-absorption now becomes $\Delta n$ = $\pm$1 depending on the circular polarization.




\end{document}